\documentclass[runningheads]{llncs}
\usepackage{amsmath,graphicx}
\usepackage[table,xcdraw]{xcolor}
\usepackage{booktabs}
\usepackage{arydshln}
\usepackage{caption}
\usepackage[inline,shortlabels]{enumitem}
\usepackage[colorlinks=false]{hyperref}
\usepackage{svg}
\usepackage{multirow}
\usepackage{siunitx}
\usepackage{amsmath}
\usepackage{amsfonts}
\usepackage{setspace}
\usepackage[T1]{fontenc}
\usepackage[thinc]{esdiff}

\usepackage{graphicx}
\usepackage{color}

\newlist{pipeline-enumerate}{enumerate}{1}
\setlist[pipeline-enumerate,1]{
  label=\textbf{S.\arabic*.},
  widest*=\widestnumber,
  parsep=0pt,
  itemsep=3pt
}
\newlist{inline-enumerate}{enumerate*}{1}
\setlist[inline-enumerate,1]{
  label=(\arabic*)
}
\newlist{inline-enumerate-roman}{enumerate*}{1}
\setlist[inline-enumerate-roman,1]{
  label=(\roman*)
}

\DeclareSIUnit{\pp}{\textup{p.p.}}

\begin{document}

\title{Cross-lingual Knowledge Distillation via Flow-based Voice Conversion for Robust Polyglot Text-To-Speech}
\titlerunning{Cross-lingual Knowledge Distillation via VC for Robust Polyglot TTS}
%

\author{Dariusz Piotrowski\inst{1}\orcidID{0000-0002-7974-9887}\inst{\star} \and
Renard Korzeniowski\inst{1}\orcidID{0009-0004-8759-803X}\inst{\star} \and
Alessio Falai\inst{1}\orcidID{0000-0002-6574-5607}\inst{\star} \and
Sebastian Cygert\inst{2}\orcidID{0000-0002-4763-8381}\inst{\dagger} \and
Kamil Pokora\inst{1}\orcidID{0009-0006-0756-4118} \and
Georgi Tinchev\inst{1}\orcidID{0000-0002-9910-6598} \and
Ziyao Zhang\inst{1}\orcidID{0000-0002-0617-1340} \and
Kayoko Yanagisawa\inst{1}\orcidID{0000-0002-3444-7287}}

\authorrunning{D. Piotrowski, R. Korzeniowski, A. Falai et al.}
%
\institute{Alexa AI, Amazon
\email{\{piotrod,korenard,falai,kamipoko,gtinchev,zhaziyao,yakayoko\}@amazon.com}\\ \and
Gdańsk University of Technology, Gdańsk, Poland}
\maketitle              

\begin{abstract}
In this work, we introduce a framework for cross-lingual speech synthesis, which involves an upstream Voice Conversion (VC) model and a downstream Text-To-Speech (TTS) model. The proposed framework consists of 4 stages. In the first two stages, we use a VC model to convert utterances in the target locale to the voice of the target speaker. In the third stage, the converted data is combined with the linguistic features and durations from recordings in the target language, which are then used to train a single-speaker acoustic model. Finally, the last stage entails the training of a locale-independent vocoder. Our evaluations show that the proposed paradigm outperforms state-of-the-art approaches which are based on training a large multilingual TTS model. In addition, our experiments demonstrate the robustness of our approach with different model architectures, languages, speakers and amounts of data. Moreover, our solution is especially beneficial in low-resource settings.
\end{abstract}

\section{Introduction}
\label{sec:intro}

{\let\thefootnote\relax\footnotetext{$^\star$ Equal contribution.}}
{\let\thefootnote\relax\footnotetext{$^\dag$ Work done while at Amazon.}}
\setcounter{footnote}{0}

Polyglot Text-To-Speech (TTS) systems, which rely on training data from monolingual speakers in multiple language variants (from here on referred to as \textit{locale}), enable speakers in the training corpus to speak any language present in the same training corpus \cite{polyglot1,polyglot2,sota_polyglot,data_composition_polyglot}. Such State-Of-The-Art (SOTA) models are able to achieve impressive results in cross-lingual synthesis scenarios, in terms of the high-quality naturalness and accent generated in the synthesised audios. During training, the model learns to decorrelate the embedding spaces of speaker and language conditionings, which enables inference-time-synthesis of \{\textit{speaker, language}\} pairs unseen during training. Such a training paradigm is the mainstream methodology of building Polyglot TTS systems, which is referred to as \textit{Standard Polyglot} hereinafter. One major drawback of \textit{Standard Polyglot} is that the model needs to have significant capacity to be able to disentangle speaker and language characteristics \cite{data_composition_polyglot,tts-synthetic-data-google}. This means the deployment of such models on low-resource devices is extremely challenging. Moreover, existing studies \cite{data_composition_polyglot,polyglot_representation} have revealed that this approach can be sensitive to data composition.

In this work, we address the issue of deploying large Polyglot models in  computationally-constrained settings, by proposing a framework based on a high-capacity Voice Conversion (VC) model, working in conjunction with a low-capacity, single-speaker, monolingual acoustic model. The key point of this solution is the decoupling of speaker-language disentanglement and TTS tasks, which are realised in a single model with \textit{Standard Polyglot} approaches. Specifically, we shift the demanding task of speaker-language disentanglement to the upstream VC model. In addition to its ability to deliver disentanglement, the robustness of our VC model makes it much less sensitive to the composition of the training corpus, thus easing the burden of scaling the framework to new speakers and locales.
Extensive evaluations show that, when compared to the SOTA \textit{Standard Polyglot} approach \cite{sota_polyglot}, our framework yields significantly better naturalness and accent similarity and on par speaker similarity. Moreover, the robustness of our framework is confirmed in experiments with varying model architecture, target speaker, target locale, and dataset size. To sum up, the contributions of this work are as follows: \begin{enumerate}
    \item We propose a new paradigm for cross-lingual TTS based on an upstream VC model and a downstream monolingual TTS model
    \item We demonstrate, through extensive evaluations, that it beats SOTA and is robust with regard to data composition as well as architecture type and size
    \item We demonstrate that the proposed approach is especially effective under low-resource constraints, enabling the use of lightweight architectures
\end{enumerate}

\section{Related work}
The idea of using synthetic data to support TTS tasks was explored before. Some works rely on VC as a data augmentation technique, for improving signal quality in low data regimes \cite{synth_data_1,synth_data_2} or for style transfer tasks \cite{comini22_interspeech}. Most closely related to our work, authors in \cite{tts-synthetic-data-google} presented a system which uses synthetic data for accent transfer tasks, but they have an upstream TTS model, while we explore VC techniques, and their downstream acoustic model and vocoder pair is more heavy-weighted than ours, as their focus is solely on reliability. Other approaches present so-called Accent Conversion (AC) techniques on disentangling speaker and accent \cite{accent-conv,accent-conv-ppg}, but they usually work in an in-lingual fashion (e.g. by converting a British English speaker to sound like an American English one); in contrast, this work focuses on more generic cross-lingual applications.

The idea of relying on cross-lingual VC to enable Polyglot TTS has already been investigated as well, but previous works \cite{related2,related3} mainly present results on legacy VC and TTS models, whereas we do so using modern, SOTA architectures. Moreover, we rely on mel-spectrograms alone for both VC and TTS models, instead of using a diverse set of features, such as phonetic posteriorgrams (PPG), as was done in \cite{related3,related5}. Other approaches \cite{related8} position the VC model for speaker identity conversion after the TTS model, which is not ideal for latency. In addition, some works \cite{related3,related5} specifically target bilingual scenarios with various restrictions (e.g. in \cite{related3} the system is designed for one specific target speaker), whereas our approach can be scaled to any target speaker-language combination. Solutions presented in \cite{related7} can also be scaled to any speaker-language pair, but the main differences with our work are the following:
\begin{inline-enumerate}
    \item we show robustness under varying assumptions, thus making our experimental validation more comprehensive;
    \item we specifically focus on computationally-constrained applications;
    \item we show that our approach to vocoder training works well in a locale-independent fashion, further highlighting the scalability of the proposed solution.
\end{inline-enumerate}

\section{Proposed approach} 
\subsection{Framework}
\begin{figure}[htb]
    \setlength{\belowcaptionskip}{-7pt}
    \centering
    \includegraphics[width=.75\linewidth]{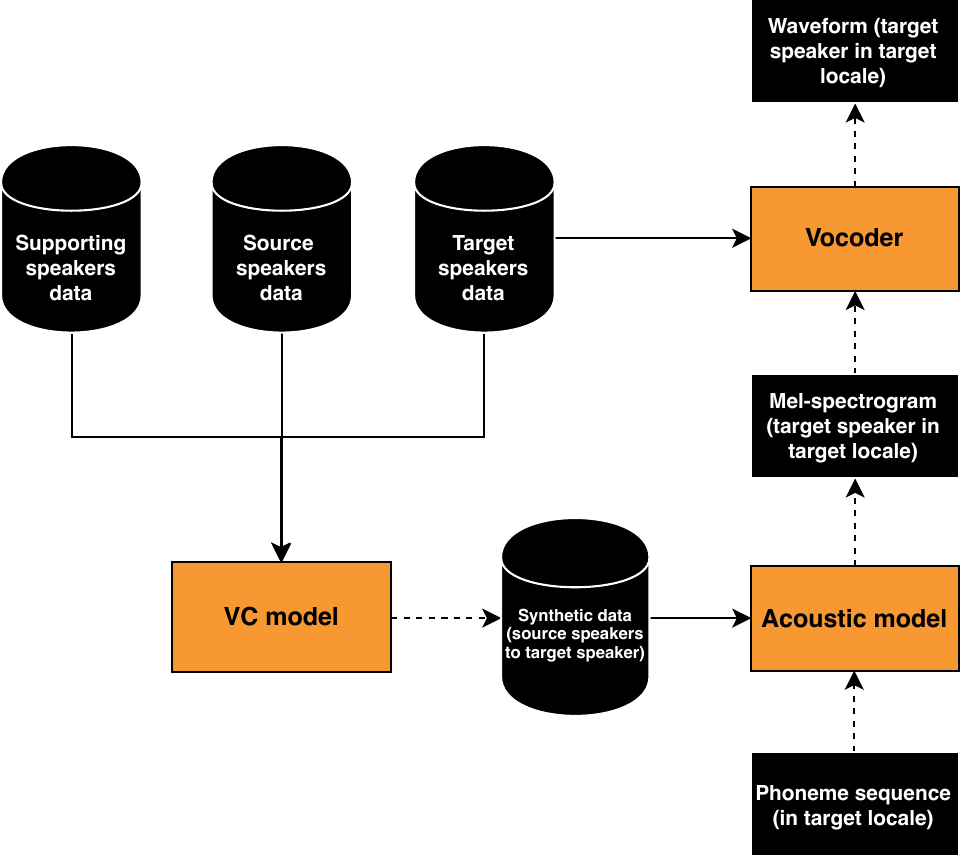}
    \caption{High-level diagram of the proposed framework, where $\rightarrow$ means training, while $\dashrightarrow$ stands for inference.}
    \label{fig:big_picture}
\end{figure}

\noindent We propose a $4$-stage approach for building a Polyglot TTS system (\textit{VC-based Polyglot}), with the goal of synthesising cross-lingual speech from a \textit{target speaker} in a \textit{target locale}. In addition, we define \textit{source speakers} as the speakers whose native locale is the \textit{target locale}; whereas the locales of \textit{supporting speakers} are not the \textit{target locale}. Our approach consists of the following steps.

\begin{enumerate}
    \item\label{pipeline-step-1} Train a many-to-many VC model, using training data from \textit{source}, \textit{target} and \textit{supporting speakers}.
    \item\label{pipeline-step-2} Convert \textit{source speakers}’ identities in the original audio files to that of the \textit{target speaker}, by using the trained VC model from \ref{pipeline-step-1}, thus creating the synthetic dataset.
    \item\label{pipeline-step-3} Use the synthetic dataset from \ref{pipeline-step-2} to train a single-speaker, monolingual acoustic model, to produce speech with \textit{target speaker} identity in the \textit{target locale}. \looseness=-1
    \item\label{pipeline-step-4} Train a speaker-specific locale-independent vocoder using original data from the \textit{target speaker}.
\end{enumerate}

When building the synthetic dataset in \ref{pipeline-step-2}, we use original phoneme durations (from \textit{source speakers}) instead of re-running forced alignment, as the latter leads to the acoustic model cutting off utterances and producing unnatural prosody. Then, at inference time we discard the upstream VC model and chain the trained acoustic model and vocoder, to output waveforms. Note that \ref{pipeline-step-2} and \ref{pipeline-step-3} can be repeated for generating synthetic data in multiple \textit{target locales}, without the need to re-train the VC model. Figure \ref{fig:big_picture} illustrates how different components interact with each other at training and inference times.

Note that the framework configuration described above is not the only possible option. For example, in \ref{pipeline-step-3}, we empirically find that adding \textit{supporting speakers} in VC model training does not improve (nor does it degrade) synthesis quality in the \textit{target locale} for downstream models. Also, in \ref{pipeline-step-4}, early results show that using \textit{source speakers} alongside original data from the \textit{target speaker} is only beneficial for some \textit{target locales}. Moreover, an alternative solution for \ref{pipeline-step-4} would be to train the vocoder in a locale-specific fashion, by generating synthetic waveforms $w$ from synthetic mel-spectrograms $m$ using a powerful universal vocoder (such as the one from \cite{universal-vocoder}), and training on $(m,w)$ pairs. We adopt the locale-independent procedure described in \ref{pipeline-step-4} because doing so renders more perceptually expressive speech.

\subsection{Model description}
\label{subsec:models}
For our upstream VC model, we initially considered normalising flows \cite{springvc,springvc2} and Variational Auto-Encoder (VAE) \cite{Karlapati2022} based architectures. However, given that the latter is particularly sensitive to the dimension of the auto-encoder bottleneck \cite{autovc}, which clashes with our requirement for robustness, we decided to adopt the flow-based VC approach. The main reason for selecting an upstream VC model instead of a TTS one is that the former operates on speech-to-speech mappings, which are easier to learn than the more ill-defined text-to-speech ones. This, in turn, leads to more natural synthetic data than their TTS counterparts \cite{springvc}.

\begin{figure}[tb]
    \setlength{\belowcaptionskip}{-11pt}
    \centering
    \includegraphics[width=0.85\linewidth]{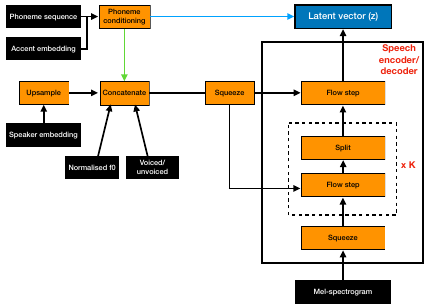}
    \caption{The many-to-many non-parallel VC model used to generate synthetic datasets. The green line denotes phoneme conditioning, while the blue line the phoneme-based prior.}
    \label{fig:spring_vc}
\end{figure}

In particular, the VC model we rely on is the text-conditioned non-parallel many-to-many VC model with fixed prior described in \cite{springvc} and depicted in Figure \ref{fig:spring_vc}. The VC model topology is based on Flow-TTS \cite{flow-tts} and modified to accept F0 (interpolated and normalised at the utterance level) and V/UV (binary voiced/unvoiced flag) conditionings. Phoneme embeddings are also enriched with accent one-hot encodings, while speaker embeddings are extracted using a pre-trained utterance-level encoder \cite{ge2e}. At training time the model learns to maximise the likelihood 
\begin{equation}
    \log{P_X(x|c)}=\log{P_Z(z|c)}+\log {\left\vert\text{det}\diffp{z}{x}\right\vert}
\end{equation} of the prior distribution \begin{equation}
    c\sim N(\mu_p,\sigma_p), \mu_p=0, \sigma_p=1
\end{equation} 
for all frames regardless of the speech content, where $z$ is the encoded latent vector and $x$ is the input mel-spectrogram. As done in \cite{springvc}, we rely on pre-trained phoneme alignments instead of using an attention mechanism.

As our main acoustic model we chose FastSpeech 2 (FS2) \cite{FastSpeech2}, which consists of feed-forward transformer-based encoder with explicit duration, energy, and F0 predictors. In order to reduce latency and model size, we remove the self-attention mechanism in the decoder, thus making it fully-convolutional. To validate the robustness of our acoustic model, we compare it with higher and lower capacity models. The former is based on gated convolutions and recurrent units (ED) \cite{edp}, while the latter is LightSpeech (LS) \cite{lightspeech}, a scaled-down version of FS2 (in two variants, as described in Section \ref{subsec:analysis}).

With low-resource embedded devices in mind, we selected MultiBand MelGAN \cite{mb_melgan}, which is able to generate high quality speech with only $0.95$ GFLOPS, as the vocoder in our pipeline. To address audio artefacts typical to GAN-based vocoders, we standardise input features in the range of $[-1, 1]$. Empirically, we find this solution equivalent to, but more suitable than, commonly used fine-tuning techniques \cite{Tacotron2,hifigan}. The practical reason for this is that no ground-truth waveforms corresponding to synthetic mel-spectrograms are available.

\section{Experiments}
\label{sec:experiments}

\subsection{Evaluation setup}
\label{subsec:evaluation-setup}
For evaluations, we use MUltiple Stimuli with Hidden Reference and Anchor (MUSHRA) \cite{mushra} tests with a scale of $0$ to $100$, through crowd-sourcing platforms, to perceptually assess naturalness, speaker similarity and accent similarity of our syntheses, as done in \cite{data_composition_polyglot,polyglot_representation}. We opted for MUSHRA tests, rather than more popular Mean Opinion Score (MOS) tests, due to the highly comparative nature of our work. In general, a MUSHRA test explicitly provides the listener with the same utterance synthesised with all of the systems to be compared, anchored by the upper and lower references. This form of testing therefore allows the listener to make direct comparisons when making their judgements, thus resulting in meaningful rank amongst systems.

For each evaluation, we use a test set of $200$ utterances, unseen during training. Each of the $60$ listeners evaluate a random set of $50$ utterances for each evaluation. When evaluating speaker similarity, testers are given a reference system, which consists of \textit{target speaker}'s recordings. Hence, the upper anchor is always equal to $100$. Recordings from \textit{source speakers} are instead used as the lower anchor. When evaluating naturalness and accent similarity, \textit{source speakers}' recordings are used as the upper anchor, while the lower anchor is a system based on phoneme-mapping, whereby an input from the \textit{target locale} is translated into the locale of the \textit{target speaker} using hard-coded rules for mapping phones between locales (based on linguistic proximity). The phoneme-mapped input is then fed into a Tacotron 2 \cite{Tacotron2} model trained on the \textit{target speaker}'s data.

To ensure the integrity of our evaluation results, we apply filtering methods to exclude potential cheaters, i.e., we remove submissions from listeners who scored all systems at extreme or default values for more than $5$ times. Moreover, to check the statistical significance of our test results, we use a pairwise two-sided Wilcoxon signed-rank test corrected for multiple comparisons with the Holm-Bonferroni method, as done in \cite{edp}.

Results in Tables \ref{tab:models} - \ref{tab:lightspeech} are reported in terms of average MUSHRA scores\footnote{We use boldface for the highest score per aspect if the gap between baseline and proposed system is statistically significant.}, along with the Closing The Gap ($CTG$) percentage, which summarises the improvement of the \textit{VC-based Polyglot} approach with respect to (w.r.t.) \textit{Standard Polyglot}. In particular, $CTG$ is calculated by
\begin{equation*}
  \begin{aligned}
    CTG &= \frac{n_v - n_s}{n_s} \times 100, \text{where} \\
    n_i &= 100 - \left(\frac{i - l}{u - l}\times 100\right), i\in\{s, v\},     
  \end{aligned}
\end{equation*}
with $s$ and $v$ representing the mean MUSHRA score for the \textit{Standard} and \textit{VC-based} approaches, respectively, and $l$ and $u$ corresponding to lower and upper anchors. All $CTG$ percentages and Difference in $CTG$ ($DCTG$) scores presented are for $p \le 0.05$, unless otherwise specified by $^{n.s.}$.

\subsection{Data composition}
\label{subsec:data}
Our default training dataset $D_1$ consists of $124$ speakers in $7$ locales\footnote{We use the ISO 639-1 nomenclature to denote locales.} (\texttt{fr-CA}, \texttt{en-GB}, \texttt{es-MX}, \texttt{en-US}, \texttt{es-US}, \texttt{fr-FR} and \texttt{de-DE}). This dataset totals to roughly $720$ hours of studio-quality recordings with \SI[round-precision=0]{24}{kHz} sampling rate. We extract $80$-dimensional mel-spectrograms with frame length of \SI[round-precision=2]{50}{ms} and frame shift of \SI{12.5}{ms}, which is then used to train the VC model and the \textit{Standard Polyglot} acoustic model. For the experiment on varying dataset size, we also test the proposed approach on a smaller dataset $D_2\subset D_1$, which consists of data from a subset of speakers in \texttt{fr-CA}, \texttt{es-MX} and \texttt{en-GB} and totals to $150$ hours.

\subsection{Experiment designs and results}
\label{subsec:analysis}
In this section, we describe the design and results of all experiments we run. Note that in all our evaluations, we compare our \textit{VC-based Polyglot} approach (\textit{VC-based}) to the \textit{Standard Polyglot} approach (\textit{Standard}) described in Section \ref{sec:intro}. In particular, Table \ref{tab:models} shows the evaluation results using different acoustic model architectures. Then, we further demonstrate the effectiveness of the \textit{VC-based} approach by changing the \texttt{en-US} \textit{target speaker} $S_1$ to \texttt{en-GB} speaker $S_2$ in Table \ref{tab:spks}, changing the \textit{target locale} to \texttt{fr-CA} in Table \ref{tab:langs}, and changing the dataset to $D_2$ in Table \ref{tab:sizes}. Finally, Table \ref{tab:lightspeech} shows that the \textit{VC-based} approach renders even more superior performance when downsized. As for the training setup, we use the same hyper-parameters as described in \cite{FastSpeech2,lightspeech,edp}. Note that for the \textit{VC-based Polyglot} approach, we train one single-speaker monolingual acoustic model for each \textit{target locale}.

The goal of Experiment $1$ (Table \ref{tab:models}) is to assess the quality of the \textit{VC-based} approach against the \textit{Standard} baseline. To this end, we rely on dataset $D_1$, \textit{target speaker} $S_1$, \textit{target locale} \texttt{es-MX} and FS2 as model architecture. Table \ref{tab:models} shows the superior performance of our proposed approach in all of the evaluated aspects for the FS2 architecture. Replicating the same experiment with ED as acoustic model results in the same conclusions, except for the on-par performance in speaker similarity (more discussions on this in Section \ref{subsec:discussion}).

\begin{table}[hbt!]
    \centering
    \caption{MUSHRA and $CTG$ scores for models evaluated in Experiment $1$, for dataset $D_1$, speaker $S_1$ and locale \texttt{es-MX}.}
    \label{tab:models}
    \begin{tabular}{p{1.5cm}p{2.5cm}p{2.5cm}p{2.5cm}p{2.5cm}} 
        \toprule
        \toprule
        \multirow{2}{*}{\textbf{Model}} & \multirow{2}{*}{\textbf{Approach}} & \multirow{2}{*}{\textbf{Naturalness$^\uparrow$}} & \textbf{Speaker similarity$^\uparrow$} & \textbf{Accent similarity$^\uparrow$} \\ 
        \midrule
        \midrule
        Upper &  & \num{82.60} & \num{100.00} & \num{82.25} \\ 
        \midrule
        \multirow{4}{*}{FS2} & \textit{VC-based} & {\bf 69.60} & \textbf{66.38} & \textbf{71.06} \\
            & \textit{Standard} & \num{64.08} & \num{64.82} & \num{66.26} \\ 
            \cmidrule(lr){2-2}\cmidrule(lr){3-3}\cmidrule(lr){4-4}\cmidrule(lr){5-5}
            & $CTG$ & \SI{29.8}{\percent} & \SI{4.4}{\percent} & \SI{30.0}{\percent}  \\ 
        \midrule
        \multirow{4}{*}{ED} & \textit{VC-based} & \textbf{70.21} & \num{67.86} & \textbf{70.49} \\
            & \textit{Standard} & \num{65.24} & \num{68.05} & \num{66.27} \\ 
            \cmidrule(lr){2-2}\cmidrule(lr){3-3}\cmidrule(lr){4-4}\cmidrule(lr){5-5}
            & $CTG$ & \SI{28.6}{\percent} & \SI{-0.6}{\percent}$^{n.s.}$ & \SI{26.4}{\percent} \\ 
        \midrule
        Lower & & \num{42.62} & \num{22.37} & \num{20.70} \\
        \bottomrule
        \bottomrule
    \end{tabular}
\end{table}

In Experiment $2$ (Table \ref{tab:spks}), we focus on evaluating  the robustness of the approach when changing the \textit{target speakers} ($S_1$ to $S_2$). Table \ref{tab:spks} shows that when we switch to \textit{target speaker} $S_2$, the \textit{VC-based} approach is still better than the \textit{Standard} one in all aspects other than speaker similarity.

\begin{table}[hbt!]
    \centering
    \caption{MUSHRA and $CTG$ scores for models evaluated in Experiment $2$, for dataset $D_1$, speaker $S_2$ and locale \texttt{es-MX}.}
    \label{tab:spks}
    \begin{tabular}{p{2.1cm}p{2.4cm}p{2.4cm}p{2.4cm}} 
        \toprule
        \toprule
        \multirow{2}{*}{\textbf{System}} & \multirow{2}{*}{\textbf{Naturalness$^\uparrow$}} & \textbf{Speaker similarity$^\uparrow$} & \textbf{Accent similarity$^\uparrow$} \\ 
        \midrule
        \midrule
        Upper & \num{80.24} & \num{100.00} & \num{82.48} \\ 
        \midrule
        \textit{VC-based} & \textbf{70.34} & \num{67.60} & \textbf{71.85} \\
        \textit{Standard} & \num{69.86} & \num{68.69} & \num{70.54} \\ 
        \cmidrule(lr){1-1}\cmidrule(lr){2-2}\cmidrule(lr){3-3}\cmidrule(lr){4-4}
        $CTG$ & \SI{4.5}{\percent} & \SI{-3.5}{\percent}$^{n.s.}$ & \SI{10.9}{\percent} \\ 
        \midrule
        Lower & \num{34.47} & \num{28.37} & \num{19.64} \\
        \bottomrule
        \bottomrule
    \end{tabular}
\end{table}

\newpage With results from Experiment $3$ (Table \ref{tab:langs}), we show the robustness of our approach when the \textit{target locale} changes. In particular, Table \ref{tab:langs} reveals that the \textit{VC-based} approach is still better across all aspects, when changing the \textit{target locale} to \texttt{fr-CA}.

\begin{table}[hbt!]
    \centering
    \caption{MUSHRA and $CTG$ scores for models evaluated in Experiment $3$, for dataset $D_1$, speaker $S_1$ and locale \texttt{fr-CA}}
    \label{tab:langs}
    \begin{tabular}{p{2.1cm}p{2.4cm}p{2.4cm}p{2.4cm}} 
        \toprule
        \toprule
        \multirow{2}{*}{\textbf{System}} & \multirow{2}{*}{\textbf{Naturalness$^\uparrow$}} & \textbf{Speaker similarity$^\uparrow$} & \textbf{Accent similarity$^\uparrow$} \\ 
        \midrule
        \midrule
        Upper & \num{75.76} & \num{100.00} & \num{76.52} \\ 
        \midrule
        \textit{VC-based} & \textbf{74.11} & \textbf{70.69} & \textbf{70.99} \\
        \textit{Standard} & \num{73.67} & \num{67.00} & \num{67.60} \\ 
        \cmidrule(lr){1-1}\cmidrule(lr){2-2}\cmidrule(lr){3-3}\cmidrule(lr){4-4}
        $CTG$ & \SI{24.6}{\percent} & \SI{11.2}{\percent} & \SI{38}{\percent} \\ 
        \midrule
        Lower & \num{54.89} & \num{50.11} & \num{63.71} \\
        \bottomrule
        \bottomrule
    \end{tabular}
\end{table}

Furthermore, in Table \ref{tab:sizes} we show the robustness of the proposed approach when scaling down the size of the dataset. For this purpose, we repeat the experiment but switch to a smaller dataset $D_2$. As can be seen in Table \ref{tab:sizes}, the proposed approach is preferred across the board.

\begin{table}[hbt!]
    \centering
    \caption{MUSHRA and $CTG$ scores for models evaluated in Experiment $4$, for dataset $D_2$, speaker $S_1$ and locale \texttt{fr-CA}.}
    \label{tab:sizes}
    \begin{tabular}{p{2.1cm}p{2.4cm}p{2.4cm}p{2.4cm}} 
        \toprule
        \toprule
        \multirow{2}{*}{\textbf{System}} & \multirow{2}{*}{\textbf{Naturalness$^\uparrow$}} & \textbf{Speaker similarity$^\uparrow$} & \textbf{Accent similarity$^\uparrow$} \\ 
        \midrule
        \midrule
        Upper & \num{78.64} & \num{100.00} & \num{80.27} \\ 
        \midrule
        \textit{VC-based} & \textbf{71.07} & \textbf{72.43} & \textbf{75.84} \\
        \textit{Standard} & \num{66.66} & \num{68.57} & \num{73.97} \\ 
        \cmidrule(lr){1-1}\cmidrule(lr){2-2}\cmidrule(lr){3-3}\cmidrule(lr){4-4}
        $CTG$ & \SI{36.8}{\percent} & \SI{12.3}{\percent} & \SI{29.9}{\percent} \\ 
        \midrule
        Lower & \num{61.55} & \num{44.80} & \num{44.30} \\
        \bottomrule
        \bottomrule
    \end{tabular}
\end{table}

Finally, to demonstrate the effectiveness of our proposed approach when scaling down model capacity, in Experiment $5$ we repeat the previous experiments (with both \texttt{es-MX} and \texttt{fr-CA} as \textit{target locales}), but use a ``slimmed down" version of FS2 \cite{FastSpeech2}, i.e. LS \cite{lightspeech}. To further test it in a more extreme scenario, we again downsize LS, thus producing LS-S, by reducing the size of all hidden layers from $256$ to $192$ and removing one convolutional layer from both the encoder and the decoder. Table \ref{tab:lightspeech} reports the results in terms of the Difference in $CTG$ ($DCTG$), which indicates how much more we close the gap with the smaller models w.r.t. the bigger ones. The $DCTG$ score is computed as $DCTG(CTG_s, CTG_b)=CTG_s-CTG_b$ percentage points (p.p.), where $CTG_s$ is the $CTG$ score for smaller models and $CTG_b$ is the one for bigger models. Each cell of the table indicates the $CTG$ score for a specific model in the \textit{target locale}, representing the improvements gained from using the proposed approach. Overall, we can see that the \textit{VC-based} approach outperforms the \textit{Standard} baseline by a wider margin when lightweight architectures are employed.

\begin{table}[hbt!]
    \centering
    \caption{$(D)CTG$ scores in Experiment $5$ for $4$ MUSHRA tests, dataset $D_1$, speaker $S_1$ and locales \texttt{es-MX} and \texttt{fr-CA}.}
    \label{tab:lightspeech}
    \begin{tabular}{p{2.1cm}p{2.1cm}p{2.4cm}p{2.4cm}p{2.4cm}}
        \toprule
        \toprule
        \multirow{2}{*}{\textbf{Locale}} & \multirow{2}{*}{\textbf{Model}} & \multirow{2}{*}{\textbf{Naturalness$^\uparrow$}} & \textbf{Speaker similarity$^\uparrow$} & \textbf{Accent similarity$^\uparrow$} \\ 
        \midrule
        \midrule
        \multirow{4}{*}{\texttt{es-MX}} & FS2 & \SI{32.87}{\percent} & \SI{-2.71}{\percent} & \SI{25.42}{\percent} \\
                               & LS & \SI{42.27}{\percent} & \SI{-1.91}{\percent} & \SI{41.24}{\percent} \\ \cmidrule{2-5}
                               & $DCTG$ & \SI{9.40}{\pp} & \SI{0.80}{\pp}$^{n.s.}$ & \SI{15.82}{\pp} \\ \midrule
        \multirow{4}{*}{\texttt{fr-CA}} & FS2 & \SI{28.89}{\percent} & \SI{4.37}{\percent} & \SI{16.42}{\percent} \\
                               & LS & \SI{28.23}{\percent} & \SI{4.98}{\percent} & \SI{45.19}{\percent} \\ \cmidrule{2-5}
                               & $DCTG$ & \SI{-0.66}{\pp} & \SI{0.61}{\pp} & \SI{28.77}{\pp} \\ \midrule
        \multirow{4}{*}{\texttt{es-MX}} & FS2 & \SI{27.55}{\percent} & \SI{-0.98}{\percent} & \SI{30.90}{\percent} \\
                               & LS-S & \SI{38.33}{\percent} & \SI{-0.23}{\percent} & \SI{35.08}{\percent} \\ \cmidrule{2-5}
                               & $DCTG$ & \SI{10.78}{\pp} & \SI{0.75}{\pp}$^{n.s.}$ & \SI{4.18}{\pp} \\ \midrule
        \multirow{4}{*}{\texttt{fr-CA}} & FS2 & \SI{30.49}{\percent} & \SI{-0.44}{\percent} & \SI{25.97}{\percent} \\
                               & LS-S & \SI{39.49}{\percent} & \SI{-0.64}{\percent} & \SI{35.54}{\percent} \\ \cmidrule{2-5}
                               & $DCTG$ & \SI{9.00}{\pp} & \SI{-0.20}{\pp}$^{n.s.}$ & \SI{9.57}{\pp} \\
        \bottomrule
        \bottomrule
    \end{tabular}
\end{table}

\subsection{Discussion}
\label{subsec:discussion}
As presented in Section \ref{subsec:analysis}, the proposed solution is preferred overall, while the only aspect where there is a non-statistically significant gap between \textit{VC-based Polyglot} and \textit{Standard Polyglot} (where the latter got more votes) is speaker similarity. We attribute this to the choice of the upstream VC model. In particular, although flow-based models are not widely-known to suffer from the speaker-accent trade-off, where speaker similarity degrades as accent similarity improves (and vice-versa), we believe that more powerful density-based models, such as those based on diffusion \cite{vc-diffusion}, have the potential of closing this gap in terms of speaker similarity.

We also hypothesise that the increase in naturalness scores can be attributed to the fact that in the \textit{VC-based Polyglot} models the generated prosody follows that of the source speaker(s). In particular, many prosodic features are language-specific and inheriting them from source speakers (i.e. native speakers of the target language) shows to be beneficial. Future work will focus on assessing the impact of prosody transfer in this cross-lingual scenario.

Furthermore, we observe that the proposed approach is robust to variations in data composition: when comparing the data generated using VC models trained on vastly different datasets, there is no statistically significant preference. In comparison, previous works with the \textit{Standard Polyglot} approach explicitly report how the mixture of training data has a significant impact on final performance \cite{data_composition_polyglot}.

\section{Conclusions}
\label{sec:conclusions}
In this paper, we proposed a novel paradigm for building highly robust Polyglot TTS systems. The introduction of a powerful upstream VC model successfully lifted the burden of learning disentangled speaker-language representations from the downstream acoustic model. This, in turn, allowed for effective use of high-performance architectures targeting low-powered devices. In summary, our proposed approach is able to simultaneously
\begin{inline-enumerate}
    \item outperform the current SOTA Polyglot TTS systems in terms of naturalness and accent similarity, while maintaining on-par performance in speaker similarity,
    \item improve model robustness w.r.t. training data composition and
    \item enable the use of lightweight architectures for speech synthesis on resource-constrained devices.
\end{inline-enumerate}
Such results are backed by controlled experiments using different model architectures, languages, speakers and dataset sizes. \looseness=-1

For future work, we will test how well our proposed approach scales in more diverse scenarios. In addition, we believe that by employing more advanced VC techniques, there would be further improvements in the downstream task, especially in terms of speaker similarity.

%
%
%
%
\begin{spacing}{0.9}
    \bibliographystyle{splncs04}
    \bibliography{samplepaper}
\end{spacing}

\end{document}